\documentclass[preprint, 12pt]{elsarticle}
\usepackage[left=2.54cm,top=2.54cm,right=2.54cm,nohead]{geometry}
\usepackage{natbib}
\usepackage{setspace}
\usepackage[dvips]{epsfig}
\usepackage{color}
\usepackage{amsmath}
\graphicspath{{figures/}}

\author[rvt]{Simon Bogner}
\ead{simon.bogner@informatik.uni-erlangen.de}

\author[rvt]{Ulrich R\"ude}
\ead{ruede@informatik.uni-erlangen.de}

\address[rvt]{Lehrstuhl f\"ur Systemsimulation, 
	Universit\"at Erlangen-N\"urnberg,
	Cauerstra{\ss}e 6,
	91058 Erlangen}

\title{Liquid-gas-solid flows with lattice Boltzmann: Simulation of floating bodies}

\begin{document}

\begin{abstract}
  This paper presents a model for the simulation of liquid-gas-solid flows by means of the
  lattice Boltzmann method. The approach is built upon previous works for the simulation
  of liquid-solid particle suspensions on the one hand, and on a liquid-gas free surface
  model on the other. We show how the two approaches can be unified by a novel set of
  dynamic cell conversion rules.  For evaluation, we concentrate on the rotational
  stability of non-spherical rigid bodies floating on a plane water surface -- a classical
  hydrostatic problem known from naval architecture. We show the consistency of our method
  in this kind of flows and obtain convergence towards the ideal solution for the measured
  heeling stability of a floating box.
\end{abstract}

\maketitle

\section{Introduction}

Since its establishment the lattice Boltzmann method (LBM) has become a popular
alternative in the field of complex flow simulations \cite{Succi}. Its application to
particle suspensions has been propelled to a significant part by the works of Ladd et
al. \cite{Ladd1993,Ladd2007} and Aidun et al. \cite{AidunEtAl1,AidunEtAl2,ALD1998}. Based
on the approach of the so-called \emph{momentum exchange method}, it is possible to
calculate the hydromechanical stresses on the surface of fully resolved solid particles
directly from the lattice Boltzmann boundary treatment. In this paper, the aforementioned
fluid-solid coupling approach is extended to liquid-gas free surface flows, i.e., the
problem of solid bodies moving freely within a flow of two immiscible fluids. We use the
free surface model of \cite{KoernerEtAl,Pohl2007} to simulate a liquid phase in
interaction with a gas by means of a \emph{volume of fluid} approach and a special
kinematic free surface boundary condition. I.e., the interface of the two phases is
assumed sharp enough to be modeled by a locally defined boundary layer. This boundary
layer is updated dynamically according to the liquid advection by a set of cell conversion
rules.

This paper proposes a unification of the update rules of the free surface model with those
of the particulate flow model, which also requires a dynamical mapping of the respective
solid boundaries to the lattice Boltzmann grid. As described in \cite{Bogner09}, the
resulting scheme allows full freedom of motion of the solid bodies in the flow, which can
be calculated according to rigid body physics as in \cite{IglbergerEtAl}. We demonstrate
the consistency of the combined liquid-gas-solid method by means of a simple advection
test with a floating body in a stratified liquid-gas channel flow, and discuss the main
source of error in the dynamic boundary handling with particles in motion.


We further apply our method to the problem of rotational stability of rigid floating
structures. This kind of hydromechanical problems typically emerge in marine engineering,
where the floating stability of offshore structures is of concern \cite{Massie,Derrett}, such as
the stability of a ship in a heeled position.  Because of the static nature of the addressed
problems, numerical issues arising from hydro-dynamic effects can be widely discarded,
which makes them well-suited for the verification of the involved force calculations. In
addition to that they provide a possibility to check the convergence of the simulated
liquid-gas-solid systems into a state of equilibrium.  We succeeded in showing that basic
convergence is given, provided that adequate spatial resolutions are chosen. For the
special problem of the floating stability of cuboid structures, convergence of numerical
simulations towards the analytical model was obtained.

The idea of evaluating the simulated floating stability of rigid bodies was inspired by
\cite{Fekken}, who proposed it as a test case for a Navier-Stokes based simulator
originally developed for the estimation of ``green water'' loads on ship decks
\cite{FekkenEtAl1999}. As mentioned above, lattice Boltzmann based fluid-structure
interaction techniques have been developed for the simulation of particulate flows. We
were not able to find publications discussing the application of LBM-based free surface
flows in interaction with floating structures. To our knowledge, the only approach to
handle similiar problems is the one proposed by Jan{\ss}en \cite{Janssen2010}.

\section{Method}


\subsection{Isothermal D3Q19 Lattice BGK Method}

\begin{figure}
  \centering
  \caption{D3Q19 stencil. The weights are $w_0 = 1/3$ for C, $w_1, .., w_6 = 1/18$ for W,
    E, N, S, T, B, and $w_7, .. , w_{18} = 1/36$ for TW, TE, TN, TS, NW, NE, SW, SE, BW,
    BE, BN, BS.}
  \label{fig:d3q19}
  \includegraphics[scale=0.5]{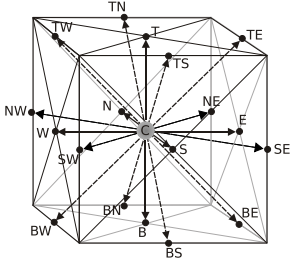}
\end{figure}

We assume the \emph{D3Q19} lattice model for 3-dimensional flows \cite{Mei3D}, with a set
of $N=19$ discrete \emph{lattice velocities} $c_i$ ($i = 0, .., N-1$). For the theoretical
considerations in this section, however, we will often fall back implicitly to the native
\emph{D2Q9} model, as it simplifies explanations and figures if they are 2-dimensional.
The lattice velocities $\vec c_i$ (also called \emph{lattice directions} or \emph{lattice
  links}) with their respective weights $\vec w_i$ ($i=0, .., N-1$), as shown in Fig
\ref{fig:d3q19}, are

\begin{equation}
  \vec{c_i} =
  \begin{cases} 
    (0,0,0) \\
    (\pm 1, 0, 0), (0, \pm 1, 0), (0, 0, \pm 1) \\
    (\pm 1, \pm 1, 0), (0, \pm 1, \pm 1), (\pm 1, 0, \pm 1) .
  \end{cases}
  \label{eq:links}
\end{equation}
In the following $\overline{i}$ is used to denote the index of the lattice velocity
$\vec{c}_{\overline{i}}$ with $\vec{c}_{\overline{i}} = - \vec{c}_i$.



Let $s_x$, $s_y$, $s_z$ be positive real numbers divisible by the \emph{spatial
  resolution}, constant $\delta_x$.  The domain $[0, s_x] \times [0, s_y] \times [0, s_z]$
is divided into \emph{cells}, i.e., cubic volumes of length $\delta_x$, which yields a
computation domain of $\frac{s_x}{\delta_x} \times \frac{s_y}{\delta_x} \times
\frac{s_z}{\delta_x}$ discrete lattice cells. Spatial quantities like $s_x$, $s_y$, $s_z$
and $\delta_x$ are commonly given in a certain unit of length (e.g., metres). However --
when dealing with LBM specific computations -- dimensionless lattice coordinates are used:
Spatial coordinates are thus given in the following as multiples of $\delta_x$. By
speaking of a cell $(x,y,z)$, where $x$, $y$ and $z$ are positive integer numbers we mean
the lattice cell with respective volume $[x, x+1] \times [y, y+1] \times [z,z+1]$ in the
lattice. We refer to the point $(x+0.5, y+0.5,z+0.5)$ as the \emph{cell center}. For each
lattice direction $i = 0, .., N-1$ we name $f_i(\vec{x}, t)$ the \emph{particle
  distribution function} (PDF) of the direction $c_i$ in cell $\vec{x}$ and of time step
$t$.

The lattice BGK propagation scheme can be derived from the classic Boltzmann equation with
the collision operator substituted by the \emph{Bhatnagar – Gross – Krook} (BGK) operator \cite{Luo2000}. Including an external
force term $F_i$, the lattice BGK (LBGK) equation reads

\begin{equation}
  f_i(\vec{x} + \delta_t \vec{c_i}, t + \delta_t) = f_i(\vec{x},t) - \frac{1}{\tau}\left[ f_i( \vec{x},t) + f_{eq,i} \left( \rho(\vec{x}, t), \vec{u}(\vec{x}, t) \right)  \right] - \delta_t F_i.
  \label{eq:lbgk}
\end{equation}
$\tau$ is the dimensionless relaxation time and related to the \emph{kinematic viscosity}
$\nu$ by
$$ \tau = \frac{\nu + 1/2 c_s^2 \delta_t}{c_s^2 \delta_t}. $$ The lattice \emph{speed of
  sound} $c_s$ is a model-dependent constant. For the D3Q19 model it is $c_s =
1/\sqrt{3}$. The equilibrium function is therewith given as a so-called low Mach number
expansion of the Maxwell distribution function \cite{HeLuo1997},

\begin{equation}
  f_{eq,i}( \rho, \vec{u}) = \rho w_i \left[ 1 + \frac{ \vec{c_i}^T \vec{u} }{c_s^2} + \frac{ (\vec{c_i}^T \vec{u})^2 }{2 c_s^4} - \frac{ \vec{u}^T \vec{u} }{2 c_s^2} \right],
  \label{eq:equilibrium}
\end{equation}

\noindent and is valid only for small flow velocities, where the following constraint holds:

\begin{equation}
  Ma := \frac{ \vec{u}^T \vec{u} }{c_s^2} \ll 1.
\end{equation}

\noindent The external force term $F_i$ is used to represent gravitation (expressed as
acceleration $\vec{a}$) in the simulation. It is given by \cite{Luo1998}

\begin{equation}
  F_i = w_i \rho \left( \frac{\vec{c_i}-\vec{u}}{c_s^2} + \frac{\vec{c_i}^T \vec{u}}{2 c_s^4} \right) \vec{a}.
  \label{eq:forceterm}
\end{equation}

\noindent The local macroscopic quantities, density $\rho$ and fluid momentum $\rho \vec{u}$, are
obtained as moments from the PDFs:
\begin{equation}
  \rho = \sum_{i=0}^{N-1}{f_i},
  \label{eq:density}
\end{equation}

\begin{equation}
  \rho \vec{u} = \sum_{i=0}^{N-1}{\vec{c_i} f_i}.
  \label{eq:momentum}
\end{equation}

\begin{equation}
  \rho \vec{u} \vec{u}^T + p \mathbf{1} + S = \sum_{i=0}^{N-1}{\vec{c_i} \vec{c_i}^T f_i}
  \label{eq:momentumfluxstress}
\end{equation}

The last equation contains the momentum flux tensor $\rho \vec{u} \vec{u}^T$ and the stress
tensor, where $S$ represents the shear stresses of the flow. For a system in equlibrium,
i.e. $f_i = f_{eq,i}$ ($i=0..N-1$), the stress tensor $S$ vanishes.

\noindent Within this lattice Boltzmann model, the pressure is linearly related to the
density, by the equation

\begin{equation}
  p = c_s^2 \rho.
  \label{eq:pressure}
\end{equation}

When a simulation includes an external force, e.g. in order to simulate the effect of
gravity, stability concerns limit the density gradient and therefore the hydrostatic
pressure gradient. Buick and Greated \cite{BuickGreated2000} give the following
incompressibility condition which relates the external force $\vec{a}$ to the projected
height of the simulation domain in the force direction.

\begin{equation}
  (l_x, l_y, l_z) \cdot \vec{a} \ll c_s^2.
\end{equation}
$l_x = \frac{s_x}{\delta_x}$, $l_y=\frac{s_y}{\delta_x}$ and $l_z = \frac{s_z}{\delta_x}$
are the respective numbers of fluid cells cells in the $x$- , $y$- and $z$- direction.

In practice, the calculation according to the LBGK equation (\ref{eq:lbgk}) is split up
into two steps. First, a propagation step, the \emph{stream step}, which propagates the
local PDFs of each cell along the corresponding lattice link into the neighboring
cell. The central one with zero lattice velocity remains on the same cell.
\begin{equation}
  f_i'(\vec{x}+\vec{c_i},t+1) = f_i(\vec{x},t)
  \label{eq:stream}
\end{equation} 
Now, in the \emph{collide step}, the relaxation towards the local equilibrium is performed for
each cell with the set $f_i'$, $i=0, .., N-1$ of PDFs. The external force $F_i$ is
added. This yields the PDFs of the succeeding time step.
\begin{equation}
  f_i(\vec{x}, t+1) = f_i'(\vec{x},t+1) - \frac{1}{\tau} \left[ f_i'(\vec{x}, t+1) - f_{eq,i}(\rho, \vec{u}) \right] - F_i.
\end{equation}
For a step by step derivation of the lattice Boltzmann method see \cite{Haenel}.

\subsection{Previous Work: Free Surface LBM}
\label{sec:fslbm}

\begin{figure}
  \caption{2D-Representation of a free liquid-gas boundary by interface cells. The real interface (dashed line) is captured by assigning the interface cells their liquid fraction.}
  \label{fig:interface}
  \centering
  \includegraphics[scale=0.4]{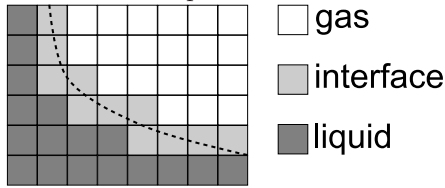}
\end{figure}

The main characteristic of the free surface model given by \cite{KoernerEtAl} is that the
dynamics of the second phase are neglected. It assumes a system of two immiscible fluids,
in which the dynamics of the first phase govern the flow completely. We refer to this
first phase as the \emph{liquid} phase and to the second phase as the \emph{gas}
phase. The layer where the two phases come in contact - the free surface - is assumed to
be thin in relation to the spatial resolution $\delta_x$. The separate regions of liquid
or gas are on the other hand assumed to be large in comparison to $\delta_x$. With this
assumption, one can simulate the liquid phase by modeling the free surface as a special
boundary condition. By means of a \emph{volume of fluid} approach that introduces a dynamic
\emph{fill level} $\phi(\vec{x}, t)$ for each lattice cell, partial filling is allowed for
cells at the free surface boundary. The fill level is the fraction of cell volume
$\delta_x^3$ that is currently filled with liquid, such that the liquid mass in a cell may
be calculated as
$$ m(\vec{x}, t) = \phi(\vec{x}, t) \cdot \rho(\vec{x}, t) \delta_x^3.$$ Cells that have a
liquid fraction $\phi<1$ and $\phi>0$ thus form a boundary for the fluid referred to as
the \emph{interface}, and need special treatment, which is outlined in the
following. We refer to those cells as \emph{interface} cells. Fig. \ref{fig:interface}
shows the interface layer between liquid and gas for a virtual free surface, drawn as a
dashed line. During the stream step, the fill level may change due to the exchange of PDFs
with neighboring cells. The mass balance between two neighboring cells, $\vec{x}$ and
$\vec{x}+\vec{c_i}$, is given by
\begin{equation}
  \Delta m_i =   
  \begin{cases} 
    0 & \text{if  $\vec{x}+\vec{c_i}$ is gas,}\\
    f_{\bar{i}}(\vec{x}+\vec{c_i}, t) - f_i(\vec{x}, t) & \text{if $\vec{x}+\vec{c_i}$ is liquid,}\\
    \frac{1}{2} \left( \phi(\vec{x}, t) + \phi(\vec{x}+\vec{c_i}, t) \right) \cdot \left( f_{\bar{i}}(\vec{x}+\vec{c_i}, t) - f_i(\vec{x}, t)  \right) & \text{if $\vec{x}+\vec{c_i}$ is interface}.
  \end{cases}
\end{equation}


The interface must further fulfill the characteristic of a free surface boundary. Because
the gas flow is neglected in the model, only the pressure of the gas is taken into account
at the interface between the two phases. Apart from that, the interface must freely follow
the liquid flow, which is provided by a special \emph{free surface boundary
  condition}. Assuming that an approximated normal vector $\vec{n}(\vec{x},t)$ of the free
surface is known for the interface cell at position $\vec{x}$, the PDFs $f_i$ with
$\vec{c_i}^T \vec{n} \leq 0$ (pointing towards the liquid phase) are set to
\begin{equation}
  f_i'(\vec{x}, t+1) = f_{eq,i}(\rho_G(\vec{x}), \vec{u}(\vec{x})) + f_{eq,\bar{i}}(\rho_G(\vec{x}), \vec{u}(\vec{x})) - f_{\bar{i}}'(\vec{x}, t+1)
  \label{eq:reconstruct}
\end{equation}
during the stream step. Here $\rho_G(x)$ is chosen such that by Eq. \ref{eq:pressure} it
matches the pressure $p_G$ of the gas phase, which must be given. Indeed, K{\"o}rner et
al. \cite{KoernerEtAl} show that this results in a zero strain rate tensor for the
boundary. This can be seen by substituting the reconstructed PDFs from Eq.
\ref{eq:reconstruct} into the formula for the momentum flux and stress tensor, i.e., the
second order moment (Eq. \ref{eq:momentumfluxstress}) of the PDFs:
\begin{align*}
\rho \vec{u} \vec{u}^T + p \mathbf{1} + S =& \sum_{i, \vec{c_i}^T \vec{n} > 0} \vec{c_i} \vec{c_i}^T f_i'(\vec{x}, t+1) + \\
& \sum_{i,\, \vec{c_i}^T \vec{n} \leq 0} \vec{c_i} \vec{c_i}^T \left[ f_{eq,i}(\rho_G, \vec{u}) + f_{eq,\bar{i}}(\rho_G, \vec{u}) - f_{\bar{i}}'(\vec{x}, t+1) \right]\\
=&  \sum_{i,\, \vec{c_i}^T \vec{n} \leq 0} \vec{c_i} \vec{c_i}^T \left[ f_{eq,i}(\rho_G, \vec{u}) + f_{eq,\bar{i}}(\rho_G, \vec{u})  \right]\\
=&  \sum_{i=0}^{N-1} \vec{c_i} \vec{c_i}^T f_{eq,i}(\rho_G, \vec{u}).
\end{align*}
Because the last sum resembles a system in equilibrium, it follows $S = \mathbf{0}$ and $p=c_s^2\rho_G=p_G$.

The normal vector $\vec{n}(\vec{x})$ is obtained as approximation from the gradient of
fill levels within a local neighborhood of the cell. It is also possible to include the
effect of surface tension directly in the boundary condition above. If a constant surface
tension parameter $\sigma$ is given and the local curvature of the interface
$\kappa(\vec{x},t)$ is known, the gas pressure is augmented by the \emph{Laplace pressure}
to
$$p_{G}' = p_{G} + 2 \sigma \kappa.$$ 

Pohl \cite{Pohl2007} proposed a method for 3D-calculations that include surface
tension. His approach is, to extract the local curvature $\kappa$ from the surface normals in a
local neighborhood as a second step after the normal computation.

In order to allow free advection of the free boundary between gas and liquid, the state of
the interface cells must be tracked carefully.  Since the interface cells represent a
boundary for the LBM scheme, it is necessary to assure a closed layer of interface cells
around the liquid cells. This is facilitated by a set of \emph{cell conversion
  rules}. Thus, no direct state transition between gas and liquid is allowed. If an
interface cell reaches a fill level $\phi(\vec{x}) \leq -\epsilon$ or $\phi(\vec{x}) \geq
1+\epsilon$ it is converted into a gas cell or a liquid cell, respectively. Since such a
conversion could introduce holes in the interface, further cell conversions from either
gas or liquid into interface cells may be triggered. The newly generated interface cells are
initialized with a fill level $\phi=0$ or $\phi=1$, respectively. This cell conversion
scheme has to be extended for the incorporation of floating objects, as shown in
Sec. \ref{sec:LGS}.

\subsection{Previous Work: Particulate Flow LBM}
\label{sec:particles}

\begin{figure}
  \centering
  \caption{A box-shaped rigid body mapped into the lattice. With a free surface flow, four different cell types have to be distinguished: Liquid, interface, gas and obstacle.}
  \includegraphics[scale=0.4]{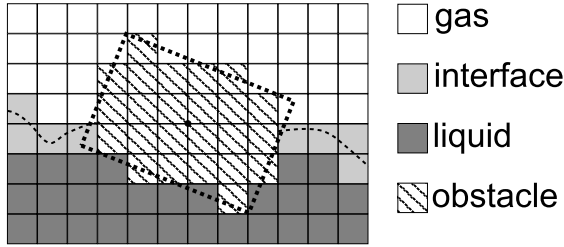}
  \label{fig:mapping}
\end{figure}

Walls and other solid obstacles that block the fluid motion must be handled with
appropriate boundary conditions.  For this paper we assume that all solid boundaries can
be modeled by a no-slip condition, i.e. the relative motion of the liquid at the boundary
is zero. This can be achieved by reflecting distribution functions, that are hitting an
obstacle cell during the stream step, to the opposed direction. If $f_{\bar{i}}(\vec{x})$
is given with an obstacle cell at $\vec{x} + \vec{c}_{\bar{i}}$ it is bounced back according to
\begin{equation}
  f'_i(\vec{x}, t+1) = f_{\bar{i}}(\vec{x},t) + \frac{2}{c_s^2} w_i \vec{c_i}^T \vec{u}_w \rho,
  \label{eq:noslip}
\end{equation}
where $\vec{u}_w$ is the velocity of the obstacle. The second term on the right hand side
accelerates the fluid according to the movement of the wall \cite{Ladd1993}.  If the solid
is not moving ($\vec{u}_w=\vec{0}$), this term will vanish, and the boundary handling is
reduced to a pure reflection.

From the boundary condition of Eq. \ref{eq:noslip} one can directly obtain the stresses
exerted on the boundary. This is known as \emph{momentum exchange}
\cite{Ladd1993,Ladd2007,MeiYuShyyLuo2002} in literature. The momentum transferred to the
wall must equal the change of momentum which results from the reflection of the PDF, since
elastic collisions are assumed. Thus, for a $f_{\bar{i}}$ reflected to $f_i'$ by
Eq. \ref{eq:noslip}, the change in momentum is
\begin{equation}
  \Delta \vec{j}_i = f_i' \vec{c_i} - f_{\bar{i}} \vec{c}_{ \bar{i}  } = 2 \vec{c_i} (f_{\bar{i}} + \frac{1}{c_s^2} w_i \vec{c_i}^T \vec{u}_w).
  \label{eq:momentumexchange}
\end{equation}
The simulation of immersed rigid bodies behaving according to Newtonian physics is
facilitated by discretizing the shape of a body to the lattice. For every time step the
no-slip boundary is set according to the object position and orientation. If the center of
a cell is a point within the volume represented by the shape of the body, it is treated as
an obstacle cell (Fig. \ref{fig:mapping}). The movement of the rigid bodies resulting from
the stresses, Eq. \ref{eq:momentumexchange}, including the resolution of body-body
collisions, is realized in practice by coupling the lattice Boltzmann algorithm to a rigid
body physics engine as in \cite{IglbergerEtAl,BinderEtAl}. The resulting body motion is
then fed back to the fluid simulation by setting the boundary condition according to
Eq. \ref{eq:noslip} in the cells covered by the obstacle. Hereby $\vec{u_w}$ is given by
the velocity of the body.

A given particle must behave according to the stresses exerted by the fluid. The net force
and torque on the particle are obtained from the momentum flux according to
Eq. \ref{eq:momentumexchange}, summed over the discretized particle surface. Let $N_o$ be
the set of grid points $\vec{x}$ next to the discretized particle surface. For a cell
$\vec{x} \in N_o$, let $I_o(\vec{x})$ be the set of indices $i \in I_o(\vec{x})$, where
$\Delta \vec{j}_i(\vec{x})$ is defined due to a reflection of PDFs as stated above. Then
the discrete surface integrals for force $\vec{F}_{net}$ and torque $\vec{T}$,
respectively, are given by

\begin{equation}
  \vec{F}_{net} = \sum_{\vec{x} \in N_o}{ \sum_{i \in I_o(\vec{x})} { \Delta \vec{j}_i(\vec{x}) \cdot \frac{\delta_x}{\delta_t} } },
  \label{eq:netforce}
\end{equation}

\begin{equation}
  \vec{T} = \sum_{\vec{x} \in N_o}{ \sum_{i \in I_o(\vec{x})}{ (\vec{x} - \vec{o}) \times \Delta \vec{j}_i(\vec{x}) \cdot \frac{\delta_x}{\delta_t} }  }.
  \label{eq:torque}
\end{equation}

\noindent In the latter sum, $\vec{o}$ is the center of gravity of the object. A non-zero torque
acting on a freely moving rigid body will induce rotational movement of the body.

The momentum exchange method has become a kind of quasi-standard for the simulation of
particle suspensions with Lattice Boltzmann and it has been mentioned in a large number of
publications with only slight modifications. The main difference here to the works of
Ladd \cite{Ladd1993, Ladd2007} or Aidun et al. \cite{AidunEtAl1,AidunEtAl2} is, that in
our case the obstacle cells are always fluid blocking nodes. In the mentioned references
there are ``virtual'' fluid nodes inside the discretized object shapes, which can be
reactivated when uncovered by from the object. This approach, however, would be difficult
to maintain in the presence of a liquid-gas flow, since it is unclear how the liquid-gas
interface would have to be treated when intersecting with the volume of a particle.

\subsection{A Liquid-Gas-Solid LBM}
\label{sec:LGS}

\begin{figure}
  \centering
  \includegraphics{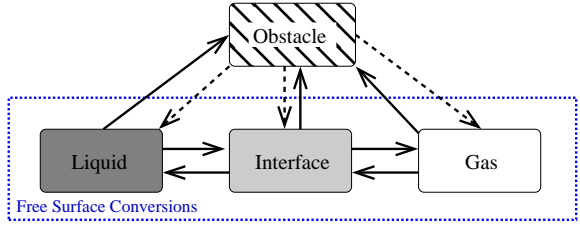}
  \caption{Allowed state transitions for cell types. With moving obstacles the number of
    transitions increases from four to ten. The free surface type conversions are framed
    by a dotted line. The dashed arrows indicate critical state changes from solid to fluid.}
  \label{fig:conversions}
\end{figure}

With the free surface model of Sec. \ref{sec:fslbm} and the additional presence of moving
obstacles from Sec \ref{sec:particles}, there are now two different boundaries for the
liquid phase that must be treated dynamically. Fig. \ref{fig:mapping} illustrates the
situation for a box floating on the free surface of a liquid, simulated according to
Sec. \ref{sec:fslbm}. Since there are now four different types of cell states to consider
(liquid, interface, gas, and solid) that may change according to fluid flow and object
movement, the need for a sophisticated cell conversion algorithm arises. In the
following, we show that these conversions can be organized in such a way that a consistent
boundary around the liquid phase is assured in every time step. 

Fig. \ref{fig:conversions} shows that the total number of conversions consists of ten
state transitions. The transitions at the bottom are those of the advection scheme of the
free surface model of Sec. \ref{sec:fslbm} and need no further examination. The remaining
transitions arise from the mapping of the solid to the grid. We assume the velocity of the
object is small enough for conversions \emph{into} obstacle to occur only at positions in
the neighborhood of other obstacle cells; while conversions \emph{from} obstacle to one of
the three fluid states (liquid, interface or gas) can occur only at positions in the
neighborhood of fluid cells. This is no additional limitation, but a direct consequence of
the low Mach number limitation of the lattice Boltzmann scheme. Consider now an object
penetrating the free surface as shown in Fig. \ref{fig:mapping}. While conversions into
obstacle cells are generally safe, the conversions from obstacle into fluid could lead to
invalid lattice configurations with holes in the interface layer. Thus the correct cell
type, i.e., liquid, interface or gas, needs to be determined such that a valid lattice
configuration is adhered for the next time step. This is achieved by an update rule based
on the local neighborhood of a given obstacle cell.

We define the non-obstacle neighborhood of a cell as $B(\vec{x}) := \{ \vec{x} + \vec{c_i}
\; | \; \vec{x}+\vec{c_i} \text{ is no obstacle} \}$. For a prior
obstacle cell $\vec{x}$ to be converted into fluid, the state is determined as follows
\cite{Bogner09}.
\begin{itemize}
  \item $B(\vec{x})$ contains no liquid: $\vec{x}$ is converted into gas.
  \item $B(\vec{x})$ contains no gas: $\vec{x}$ is converted to liquid, $\rho(\vec{x})$ is
    interpolated from $B(\vec{x})$, and the $f_i(\vec{x})$ are set to
    $f_{eq,i}(\rho(\vec{x}), \vec{u}_w)$, where $\vec{u}_w$ is the velocity of the point on the
    object's surface, which is closest to the cell center of $\vec{x}$.
  \item $B(\vec{x})$ contains liquid and gas: $\vec{x}$ is converted into
    interface. $\rho(\vec{x})$ is interpolated from the non-gas cells in $B(\vec{x})$, and
    the $f_i(\vec{x})$ are set as in the preceding case. Additionally a fill level
    $\phi(x)$ is chosen by interpolation of the fill levels of the interface cells in
    $B(\vec{x})$. It is also possible to include the gas cells with the pressure $\rho_G$
    in the interpolation of $\rho$, since pressure can be seen as a continuous quantity
    across the free surface boundary of Eq. \ref{eq:reconstruct}.
\end{itemize}

With the free surface model of Sec. \ref{sec:fslbm}, the gas phase is only taken into
account in terms of a pressure force exerted onto the free surface. This pressure forces,
however, are also acting on the surface of the solid bodies. So, in analogy to the
boundary condition for the free surface, Eq. \ref{eq:reconstruct}, we use the equilibrium
function, Eq. \ref{eq:equilibrium}, to construct PDFs $f_{eq,i}(\rho_G, \vec{0})$ for the
lattice links from gas, or interface cells to the obstacle cell of an object. From these
the momentum transferred to the body can be calculated from Eq.
\ref{eq:momentumexchange}. The value of the constructed PDF is linearly interpolated with
the liquid's distribution function $f_i$. Thus, in presence of gas in the neighborhood of an obstacle cell, the
momentum acting on the body locally is given by
\begin{equation}
  \Delta \vec{j}_i = 2 \vec{c}_i \left( \phi \cdot f_{\bar{i}} + (1-\phi) \cdot f_{eq,i}
  (\rho_G, \vec{0}) +\frac{\phi}{c_s^2} w_i \vec{c_i}^T \vec{u}_w \right).
  \label{eq:momentuminterface}
\end{equation}
For a gas cell, $\phi=0$, only the equilibrium function of Eq. \ref{eq:momentumexchange}
remains, all other parts vanish.


\section{Simulation of Floating Bodies}

In the following sections we present the results of calculations performed with the method
of Sec. \ref{sec:LGS}, i.e., various simulations of a rigid body exposed to a free surface
flow. Since many details of the method are owed to previous particulate flow simulations,
we briefly recapitulate the most relevant results in Sec. \ref{sec:dragforceval}. We then
focus on the evaluation of forces and motion of partially immersed rigid bodies,
Sec. \ref{sec:fixedbodies}, in order to prove the basic consistency of the method. When
dealing with floating objects in gravity-stratified free surface flows, the buoyancy forces
on the discretized objects turned out to be one of the main sources of error. In the last
part, Sec. \ref{sec:floatingbodies}, we apply our method to the problem of floating
stability of rigid bodies.

\subsection{Force Computations with Spherical Particles and Staircase Approximation}
\label{sec:dragforceval}
It has to be emphasized that due to the discretization of the body shapes to lattice sites
as described in the previous section, a certain \emph{a priori} error is introduced. This
is a well-known issue in LBM-based particulate flows, and there has been a lot of research
regarding the treatment of complex, curved boundaries in lattice Boltzmann over the last
decades \cite{Mei3D,AidunClausen}. These schemes allow second order spatial accuracy in
the treatment of curved boundaries. Most of them are directly applicable for moving
boundaries, such as particles. However, in the kind of flows which we are going to discuss
in the final part of the paper, interpolation based schemes like \cite{Bouzidi2001}, or
\cite{YuScheme}, which have been applied with success to particle simulations in the past,
turned out to be little useful.

\begin{figure}
\centering
        \includegraphics[scale=0.4]{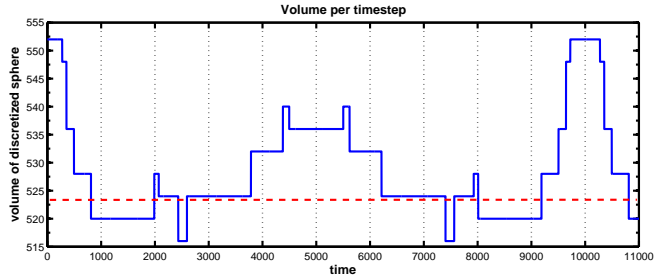} \caption{Discretized
        volume for a spherical particle traveling along an axis-aligned path with a
        velocity of $10^{-4}$. Ideal volume: $523.6$. The graph repeats periodically every
        $10000$ time steps} \label{fig:volumeerror}
\end{figure}

The solid particle model described in Sec. \ref{sec:particles} which we use throughout
this paper is identical to the one used for drag force computations on suspended particle
agglomerates in \cite{BinderEtAl,IglbergerEtAl}, where the net force deviation was studied
with spherical particles in stokes flow. The calculations were performed with various
degrees of spatial resolution and a set of different lattice relaxation parameters $\tau$
in \cite{BinderEtAl} with a sphere fixed to the center of a channel flow. The results were
found to be in good agreement with theoretical expectations. Second order boundary
conditions surpassed the staircase approximation under all tested circumstances.  In
\cite{IglbergerEtAl}, sinking sphere simulations were conducted in order to evaluate the
method with moving particles as proposed in \cite{ALD1998}. These simulations included the
influence of gravity, and used the unified boundary treatment by \cite{YuScheme} for the
solid boundaries: A particle of density $\rho_s>1$ accelerates until the frictional force
imposed by the fluid surrounding balances its weight and a terminal sinking velocity is
reached. The results showed small periodic oscillations in the measured force on a sphere
that moves with a quasi-constant velocity through the lattice. This error can be
attributed to the staircase approximation of the boundary of the
body. Fig. \ref{fig:volumeerror} shows the volume error for a sphere of radius $5\;
\delta_x$ when traveling along an axis-aligned path. The number of cells actually covered
by the sphere is depending on its exact position relative to the lattice. With the
discretized shape of a particle changing, the set of lattice links included in the
discrete surface integral around its boundary changes and so does the net force,
Eq. \ref{eq:netforce}, and torque, Eq. \ref{eq:torque}, respectively. The interpolation
based schemes applied in \cite{BinderEtAl,IglbergerEtAl} however, approximate a curved
boundary by calculating its exact distance from the cell center along the lattice link of
reflection. This information is used to correct the reflected PDFs from
Eq. \ref{eq:noslip}.  Hence, the error arising in the buoyancy force from changes in the
discrete volume of an object cannot be corrected by purely local refinements. For
liqud-solid suspensions without a free surface the problem of buoyancy force fluctuations
with fluid-blocking obstacles is usually negligible, or can be corrected by workarounds as
reported in \cite{Goetz2011}. The following subsections discuss this issue for the free
surface case.

\subsection{Forces on Buoyant Rigid Bodies}
\label{sec:fixedbodies}
Due to the staircase approximation of rigid body shapes, not only is the buoyancy force
depending on the object's position, but -- for partially immersed objects -- a staircase
function with respect to draft. This can be best explained for an axis-aligned cube of
side length $s$ in a liquid at rest with a constant water level. The buoyant force
is then given as a linear function with respect to the draft $d$:
$$F_B = \rho g s^2 d.$$ Because of the discretization error, $d$ is only exact up to the
lattice resolution $\delta_x$. This can keep buoyant objects from coming to rest at their
equilibrium position. Especially for objects with lattice-aligned planar faces such as a
cube this would sometimes cause an unphysical ``wobbling'' behavior with the object
jumping between two points slightly below and above the ideal floating position.  If on
the contrary a fixed cuboid is considered and the gauge of the liquid is varied immersing
more or less of the object's surface, the changes in the buoyant force on the object would
be continuous. This is because the volume of fluid-based interface treatment is not
limited to the spatial accuracy of the lattice. The fill level is taken into account in
the force calculations in the interface cells at the triple line of liquid, gas and solid
by Eq. \ref{eq:momentuminterface}. Test cases with a linearly rising water level around a
partially immersed, axis-aligned, fixed cuboid led to a linear increase in the buoyant
force.

\subsubsection{Free Advection Test}
\begin{figure}
        \centering \caption{Schematic view for a free advection test with a floating
        sphere in laminar free surface flow. The channel is driven by a moving wall
        boundary condition at the bottom and its longitudinal side planes. The channel
        entry and outflow are connected with each other by means of a periodic boundary
        condition.}
        \label{fig:channel}
        \includegraphics[scale=0.4]{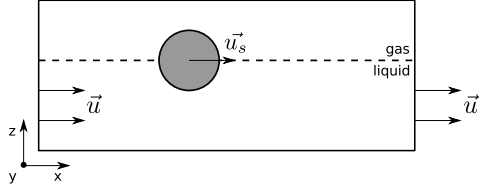}
\end{figure}

To test the consistency of the unified liquid-gas-solid algorithm, a \emph{free advection
  test} was performed in a channel flow with a constant homogeneous velocity in the
laminar regime along the x-direction, and a gravitational field along the z-axis. A
spherical particle of density $\rho_s=0.5$ was placed freely floating in the center of the
channel to be accelerated by the hydrodynamic forces acting on its
surface. Fig. \ref{fig:channel} gives an impression of the simulated setup. Because the
gas phase is neglected in the model it must not hinder the particle motion in any
way. This means, that in both cases the final velocity of the particle must equal the
fluid velocity. After the systematic drag force evaluation cited in the preceeding
section, this test might seem trivial at the first spot. But in the presence of a density
gradient and because of the changes in the discretized particle shape
(Sec. \ref{sec:dragforceval}), there are at least two important sources of error to
consider. Hence, in the following we also examine the results at critical flow velocities
that have the same order of magnitude as the expected errors.

The channel size measured $60 \times 40 \times 30$ lattice units, and the particle
diameter was $12\delta_x$. All long sides of the channel were no-slip moving walls
(Eq. \ref{eq:noslip}), in order to obtain a homogeneous velocity profile across the domain
with a constant flow velocity $u$ along the x-axis. The left ($x=0$) - and right ($x=s_x$)
- boundaries were connected periodically, by copying PDFs streamed outside at one end into
the cells at the opposing end of the domain.

The whole process of testing was performed at four different values of $\tau_{1..4}$,
namely $0.62,$ $0.8,$ $1.1,$ $1.7$ a gravitational constant of $g=10^{-5}$, and four
different channel velocities $u_{1..4} = 10^{-4},$ $5\cdot10^{-5},$ $2.5\cdot10^{-5},$
$1\cdot10^{-5}.$ In order to check for possible sources of errors the scenario was run in
a sequence of increasing complexity, starting with a simplified version of the simulation
experiment. The first series were the following.
\begin{itemize}
  \item Channel fully filled with liquid without gas; without gravitational field.
  \item Repetition without gas, but with gravitational field as stated above. The particle
    density was set to $1.0$ in this case.
\end{itemize}
In the first run the particle velocity matched the liquid velocity to the last digit
without any deviations. For the second run we were expecting errors due to oscillations in
the buoyant force as described above. These were clearly measurable, and largest for the
lowest lattice viscosity $\tau_1=0.62$, where the z-component of the object velocity
oscillated within the interval $[ -g, +g ]$. For larger $\tau$ values the effect would be damped
(e.g., by a factor of $0.1$ in the case of $\tau_4$).  There were no significant deviations
in the other velocity components of the particle.

The next test series consisted of the channel partially filled with liquid with a planar
free surface at $z=19.5$ under the influence of gravity. The particle was removed to check
the behavior of the interface cells. The free surface remained planar under all tested
circumstances, as expected. For the lowest value of $\tau=0.62$, locally spurious
interface velocities were occuring. However, the absolute deviation from the bulk
viscosity was always below $10^{-5}$, restricted locally to parts of the interfacial
layer, and therewith only critical for low velocities.

We finally tested the advection of a solid body in the free surface flow by placing the
particle with density $\rho_s=0.5$ inside the channel. The fluctuations in the lifting
force were comparable to those measured in the all-liquid domain under gravity with all
tested cases.  However, here the error also became visible in the in the other components
of the net force on the object, and thus introduced an error in the streaming velocity of
the particle. We believe that this is caused by the errors introduced by the refilling
step of the algorithm, when obstacle-to-fluid conversions occur for an object sticking
through the interface layer. As described in Sec. \ref{sec:LGS}, we initialize these cells
to a local equilibrium with parameters interpolated from a local neighborhood. This is
likely to introduce a certain error that cannot be balanced in the asymmetrical case when
an obstacle is at the boundary between gas and liquid. Even in case of liquid-solid flows
without a gas phase, the problem of initializing new liquid cells is non-trivial
\cite{Caiazzo2008}.

We have not evaluated those cases, where the empty channel showed spurious
currents. These were: $\tau_1=0.62$ with velocities $u_2,..,u_3$. In the remaining cases
the sphere would follow the current with the velocities given in
Tab. \ref{tab:velocities}. For higher flow velocities (or lower values of $g$), this error
would no longer be critical, if even apparent in the sphere velocities.

\begin{table}[ht]
\centering \caption{Errors in the velocity of the sphere for lattice gravity of
  $g=10^{-5}$: Critical if $\tau$ is low and the characteristic flow velocity of the same
  order as $g$.}
\begin{tabular}[ht]{|l|c|c|c|c|}
  \hline
  $\tau$ & $u_1$     & $u_2$  & $u_3$  &  $u_4$ \\
  \hline\hline
  0.62 & 1.1e-4 &	n.e. &	n.e.  &	n.e.   \\
  0.8 & 1.05e-4 &	5.5e-5 &	3.0e-5 &	2.0e-5  \\
  1.1 & 1.01e-4 &	5.15e-5 &	2.8e-5 &	1.3e-5 \\
  1.7 & 1.e-4 &	5.05e-5 &	2.55e-5 &	1.1e-5 \\
  \hline
\end{tabular}
\label{tab:velocities}
\end{table}

The the systematic error in the staircase approximated objects is negligible as long as
the characteristic velocity $U$ of the flow is large in comparison with the gravitational
constant. However, it should be considered in the Froude number scaling \cite{Massie} ($Fr
= U/{\sqrt{g L}},$ for a characteristic length $L$) of simulations. On the other hand
the question arises, whether a more accurate treatment of particle boundaries can
be achieved that does not suffer from the presence of a hydrostatic density gradient. We
were unable to find any literature that reports the problem explicitly. Besides increasing
the spatial resolution, or applying local grid refinement methods that entail complex
restructuring of the whole implementation of the method, we found at least one approach that
could provide a significant improvement. Based originally on Chen et al., a volumetric
interpretation \cite{Chen1998} of the lattice Boltzmann method allows the direct
calculation of the momentum flux onto finite surface elements. This can be exploited to
realize boundary conditions, like \cite{ChenEtAl1998,VerbergLadd2000,RohdeEtAl2002}, that
offer sub-grid scale accuracy. \cite{RohdeEtAl2002} presented a variant that allows an
exact description of the surface of obstacles including surface stress calculation which
is independent of surface motion. As the free surface boundary treatment,
Eq \ref{eq:reconstruct}, is in principle also a volumetric approach \cite{KoernerEtAl}, this
would yield a further unification of the method described in this paper. We expect to
explore this issue in future publications.


\subsection{Stability of Floating Bodies}
\label{sec:floatingbodies}

\begin{figure}
        \centering \caption{For both positions of the cube of specific density $\rho_s =
        1/2$ hydrostatic equilibrium is given. However the equilibrium on the left hand
        side is unstable. A stable equilibrium can be found with the cube rotated $45$
        degrees. } \label{fig:cubeExample} \includegraphics[scale=0.5]{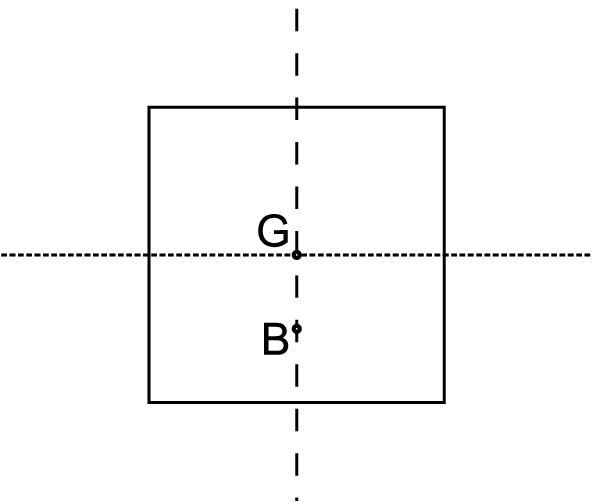} \hspace{1cm} \includegraphics[scale=0.5]{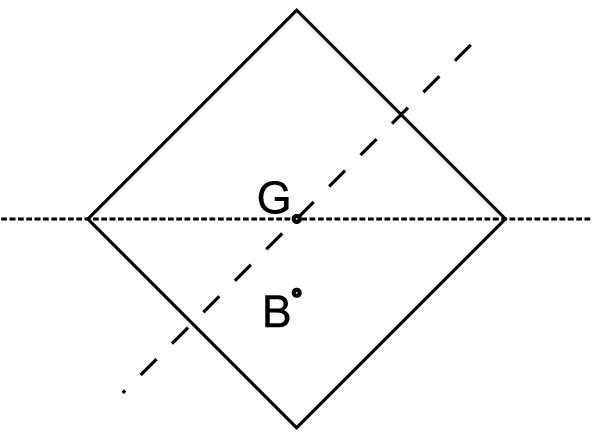}
\end{figure}

\begin{figure}
        \centering \caption{Stable floating positions for a cube of material density
        $\rho_s=1/4$ (left) and $\rho_s=3/4$ (right), respectively. Each time one corner of
        the cube is in place with the water line while another one is exactly
        half-immersed.}  \label{fig:otherStables} \includegraphics[scale=0.5]{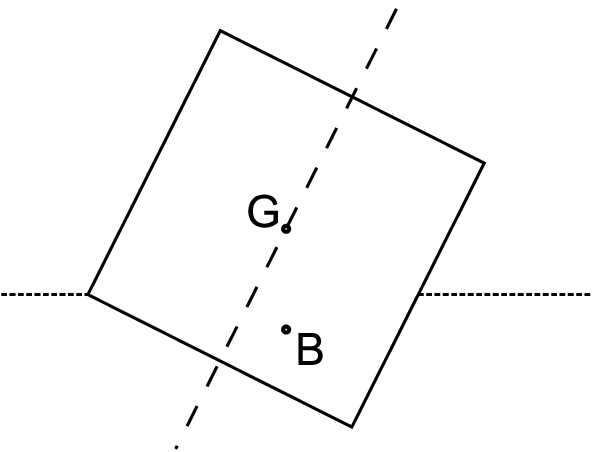} \hspace{1cm} \includegraphics[scale=0.5]{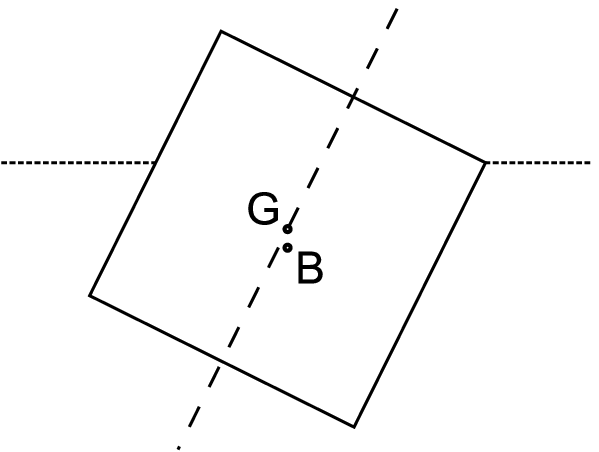}
\end{figure}

\begin{figure}                                                                                                                 
        \centering \caption{Unstable equilibrium of a cube. Because of the slight
          horizontal displacement of the center of buoyancy $B$ relative to the center of
          gravity of the cube $G$, a moment occurs that pushes the cube further from its
          upright position. The center of buoyancy $B$ is constructed as the center of
          gravity of the trapezoid, which the immersed part of the cube makes up together
          with the fluid surface. To construct $B$, the base line segment $u$ is elongated
          in one direction by the parallel opposite segment $v$, while the latter is again
          elongated in the opposite direction by base line $u$. Next, the shifted end
          points of each line segment are connected. The intersection of the connecting
          line with the line through the midpoints of $u$ and $v$ gives its center of
          gravity of the trapezoid, $B$.}
        \includegraphics[scale=0.5]{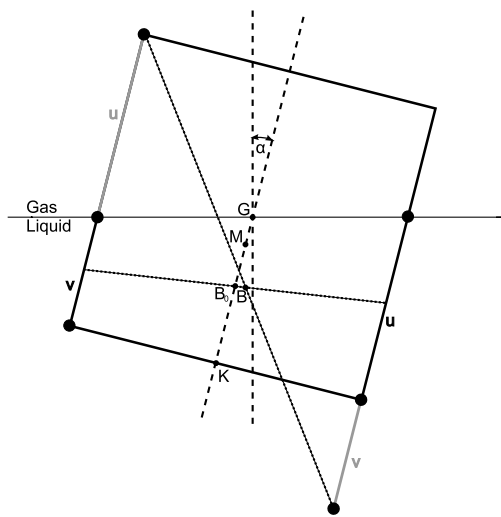} \label{fig:cubeStability}
\end{figure}

We now apply the method on the problem of floating stability of non-spherical rigid
bodies. Archimedean law appears to be quite intuitive, which states that the buoyancy
experienced by an immersed object is equal to the weight of the liquid displaced by
$\nabla$, the immersed volume part of the body (i.e., the part below the water
plane). According to that rule, a body with a specific material density $\rho_s<1.0$ will
sink down or rise up until buoyant force $\vec{F}_B = \rho g \nabla$ and gravitational
force $\vec{F}_G$ balance each other vertically. In practice, however, even if this
balance of forces is given, the resulting equilibrium is often found to be
unstable. Consider for instance a cube in 2-D as shown in Fig. \ref{fig:cubeExample}. In
both positions exactly half of the body volume is immersed and thus the body floats in
hydrostatic equilibrium.  However, it can be shown that only the right hand side with an
heel angle of $45$ degrees is stable. To see this, consider the cube slightly rotated from
its upright position, such that its central axis (dashed line in
Fig. \ref{fig:cubeExample}) is heeled by a small angle $\alpha$.  The \emph{center of
  gravity} $G$ of the cube is the point on which the gravitational force $\vec{F}_G$ acts
vertically downward. The \emph{center of buoyancy} $B$, which is defined as the center of
gravity of the immersed part $\nabla$ of the body, is slightly displaced to the left as
shown in Fig \ref{fig:cubeStability}. Thus with $\vec{F}_B$ acting on $B$ vertically
upward, a rotational moment in the heeling direction occurs, pushing the cube further from
its upright position. In the same way it can be shown that, starting from the 45 degree
rotated position (right in Fig \ref{fig:cubeExample}), small angles of heel from this
position lead to a righting moment in the opposite direction. The result is a stable
equilibrium at this position. Fig. \ref{fig:otherStables} shows two more stable floating
positions for the cube densities $\rho_s=1/4$ and $\rho_s=3/4$.

\subsubsection{Equilibrium Position of Floating Cubes}
\begin{figure}
        \centering \caption{Box rotating about its longitudinal axis (rolling motion).}
        \includegraphics[scale=0.4]{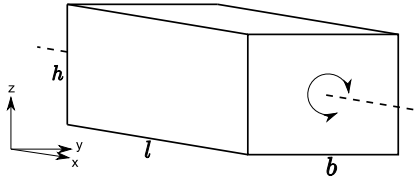}
        \label{fig:box}
\end{figure}

\begin{figure}
  \centering \caption{Box objects of density $0.75,$ $0.5$ and $0.25$, respectively,
    striving towards vertical and rotational equilibrium. The first three pictures show
    the system in motion. The last picture was taken after all visible motions had ceased.
    All snapshots were visualized from simulation output with the ray tracer
    \cite{povray}.}

\includegraphics[width=0.8\textwidth]{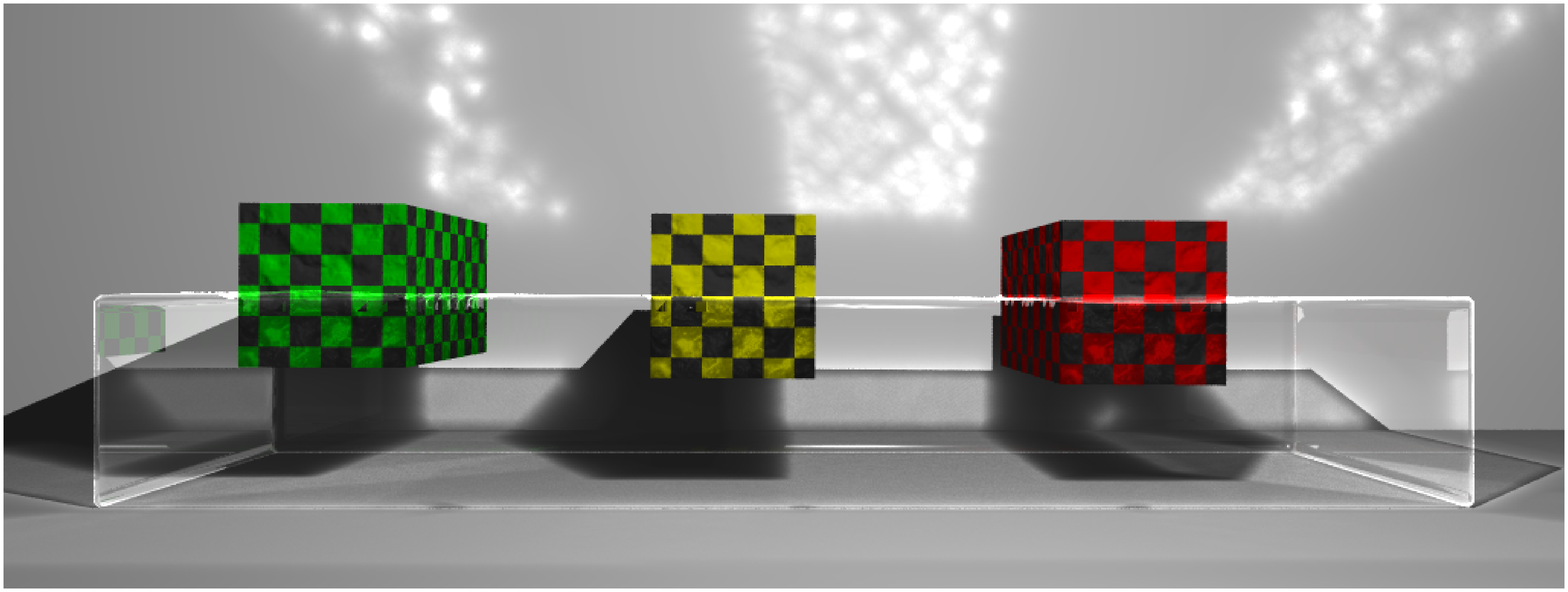}\\ 
\includegraphics[width=0.8\textwidth]{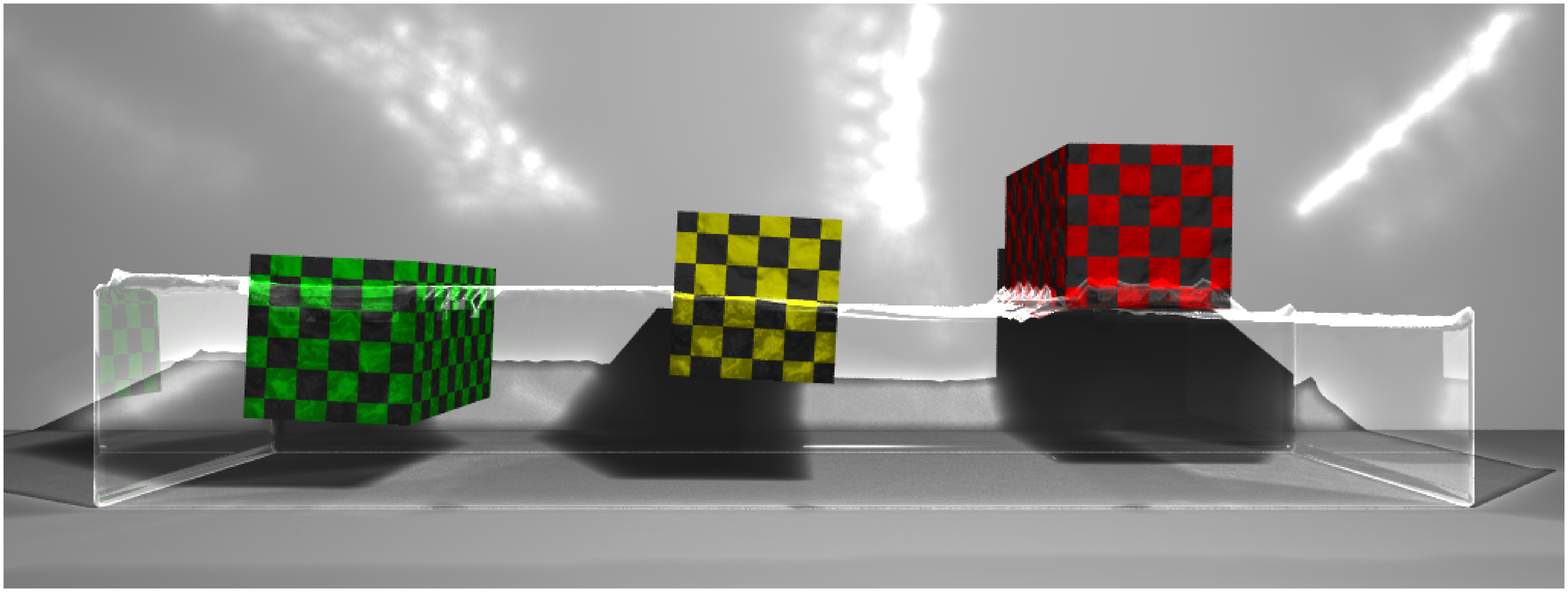}\\
\includegraphics[width=0.8\textwidth]{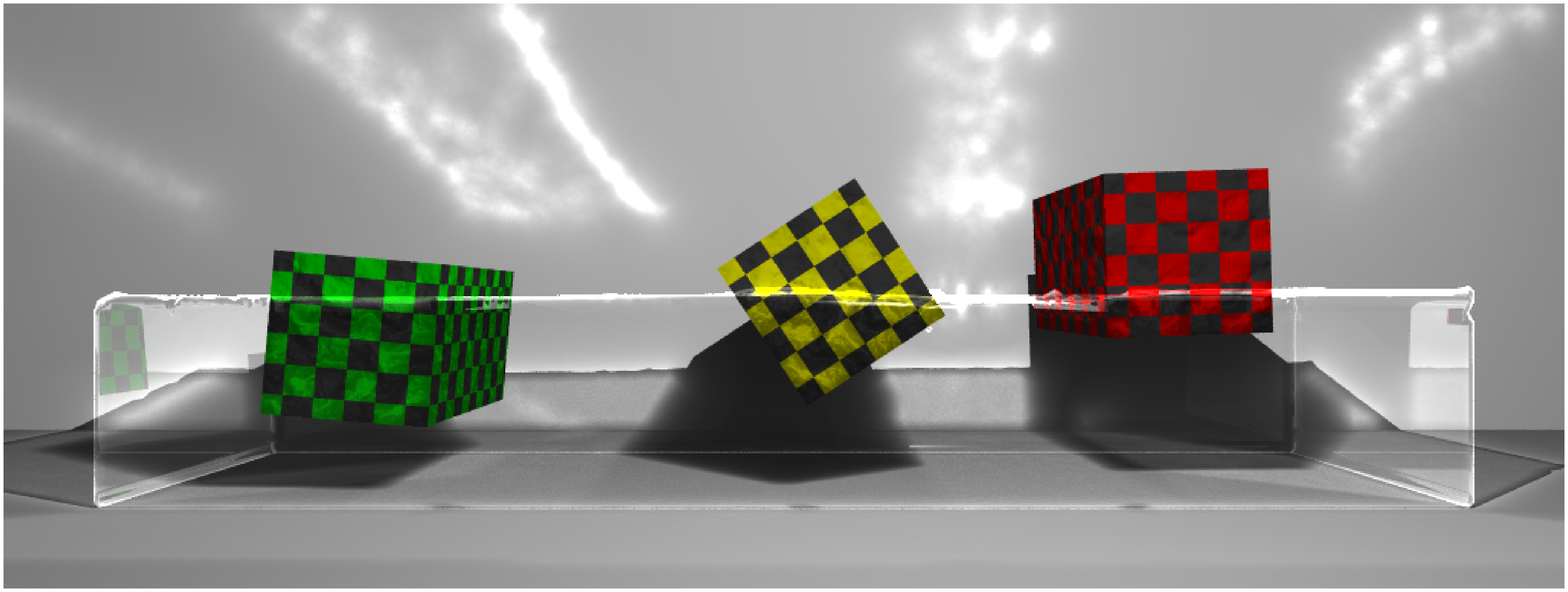}\\
\includegraphics[width=0.8\textwidth]{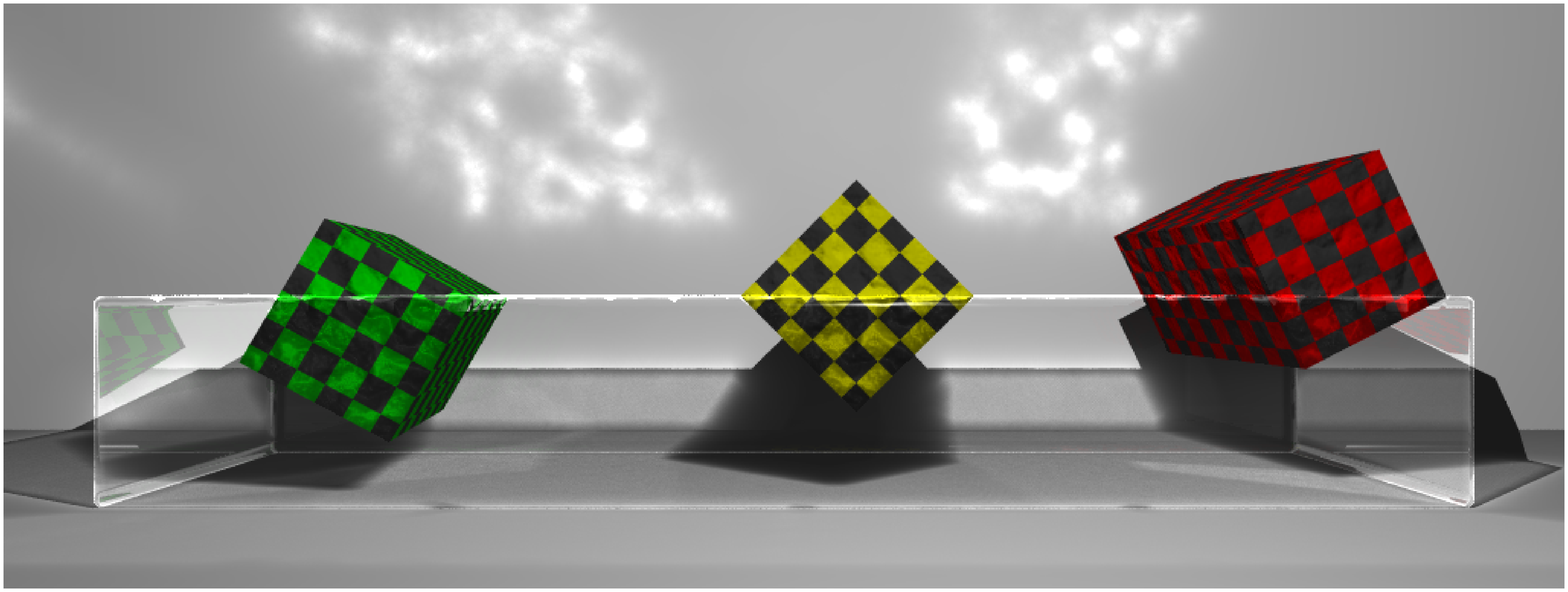}\\
\label{fig:equilibriumBoxes}
\end{figure}

Following \cite{Fekken}, we check the rotational equilibrium of floating cubes of density
$\rho_s = 0.25,$ $0.5,$ and $0.75$ in simulations. With the given densities $0.25$ and
$0.75$ a stable angle of heel is found at $26.565^{\circ}$ from the upright axis of the
cube as shown in Fig. \ref{fig:otherStables}. The cube of density $0.5$ should calibrate
to $45^{\circ}$. Since we are dealing with 3-dimensional objects, the 2D cube of side length
$s$ is represented by a cuboid of dimensions $2 s \times s \times s$, that will rotate
around its longitudinal axis only, as in the schematic of Fig. \ref{fig:box}. Inside a
domain of $130\times40\times40$ cells, the lower half was filled liquid with lattice
viscosity $\tau=1/1.9=0.526$ and initialized with hydrostatic pressure ($g=7.5 \cdot
10^{-4}$).  Three boxes with $s=16$ were placed inside the basin, slightly heeled
from their axis-aligned upright position by an angle of $2.86^{\circ}$. The simulation was
run until all fluid motions ceased and the objects came to
rest. Fig. \ref{fig:equilibriumBoxes} shows visualizations of the simulated scenario at
various time steps. Since the initial positions of the objects were far from the
equilibrium state, their motions would cause a wavy disturbance of the liquid in the
basin. It can be clearly seen from these visualizations that even at this low resolution a
strong convergence towards the ideal equilibrium was given.





\subsubsection{Righting Stability Moment of Wall-Sided Structures}

Consider again a floating structure heeled about a certain angle $\alpha$ about its
upright axis. The intersection of the line with direction $\vec{F}_B$ through $B$ with the
upright axis of a body is defined as the \emph{metacenter} $M$. If $M$ lies above the
center of gravity $G$ then the floating position of the body will be stable, otherwise
unstable (see Fig. \ref{fig:cubeStability}). This concept originates from marine
engineering and is commonly used in the characterization the floating stability of
ships and other offshore structures. A comprehensive discussion can be found
in \cite{Massie}. The lever arm of the torque arising from the horizontal displacement of
the mass centers $B$ and $G$ is given by $ \overline{GM} \sin{ \alpha }$, and hence
the \emph{righting stability moment} of the body can be calculated as
$$ m_s = \vec{F}_B \cdot \overline{GM} \sin{ \alpha }. $$ The curve $m_s(\alpha)
= \vec{F}_B \cdot \overline{GM} \sin{ \alpha }$ is called the \emph{stability curve} of
the floating object. This curve can be derived analytically for many relevant situations,
depending on the shape of the immersed structure. An important case is that of
a \emph{wall-sided} structure, where the points of the structure's surface sinking under
or rising above the water plane upon the considered angles of heel are parallel opposing
sides when the structure is upright.  If a wall-sided structure is given, then, according
to \cite{Massie,Derrett}, the Scribanti formula states
$$ \overline{B_0M} = \frac{I}{\nabla} \cdot (1 + \frac{1}{2} \tan^2 \alpha),$$ where
$B_0$ is the center of buoyancy of the upright position, $I$ is the moment of inertia of
the water plane and $\nabla$ stands for the immersed volume of the whole body. In case of
a cuboid of length $l$, width $b$, height $h$ and draft $d$ ($=\rho_s \cdot h$ --
obtained from the non-heeled position), the formula can be simplified by putting
$$\frac{I}{\nabla} = \frac{1}{12} \frac{b^2}{d},$$ and with $K$ as the keel point of the
cuboid, an expression for $\overline{GM}$ can be introduced as
$$ \overline{GM} = \overline{KB_0} + \overline{B_0M} - \overline{KG}.$$ 

\begin{figure}
        \centering \caption{Stability curves for various floating cuboids of density
          $1/2$. From the negative slope of the graph for the cube case (width:height
          ratio $4:4$) at $0^{\circ}$ angle of heel, it can be deduced that the upright
          position is unstable. The stability increases if the width is enlarged.}
        \label{fig:stabilityCurves} 
        \includegraphics[scale=0.4]{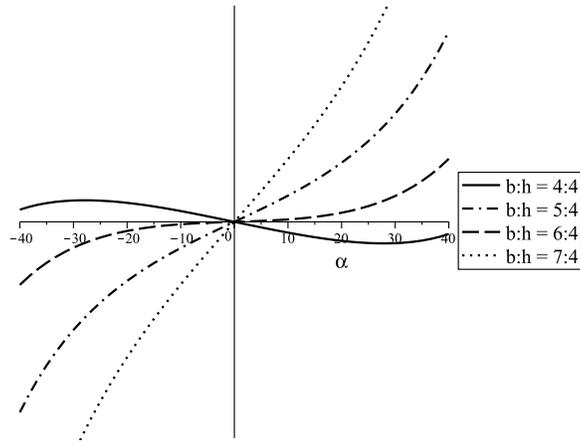}
\end{figure}

From this, the ideal stability curve of an arbitrary cuboid can be calculated.
Fig. \ref{fig:stabilityCurves} shows the schematic stability curves for some cuboid
structures of density $\rho_s=1/2$ and various width:height ratio. In case of a
cube, the stability curve has a negative slope at the origin. This shows, that the upright
position is an unstable equilibrium. Increased width yields a higher floating stability.

\begin{figure}
        \centering \caption{Stability curve of a heeled cuboid at different resolutions
        $dx = 1,$ $0.5,$ $0.25.$  The legend reads the width of the box in lattice units at the
        different resolutions.  There are significantly less deviations in the torque of
        the more stable box of width per height ratio 6:3 (a) compared to the box shape of
        ratio 5:4 (b).
        }  \label{fig:measuredMoment} \includegraphics[scale=0.35]{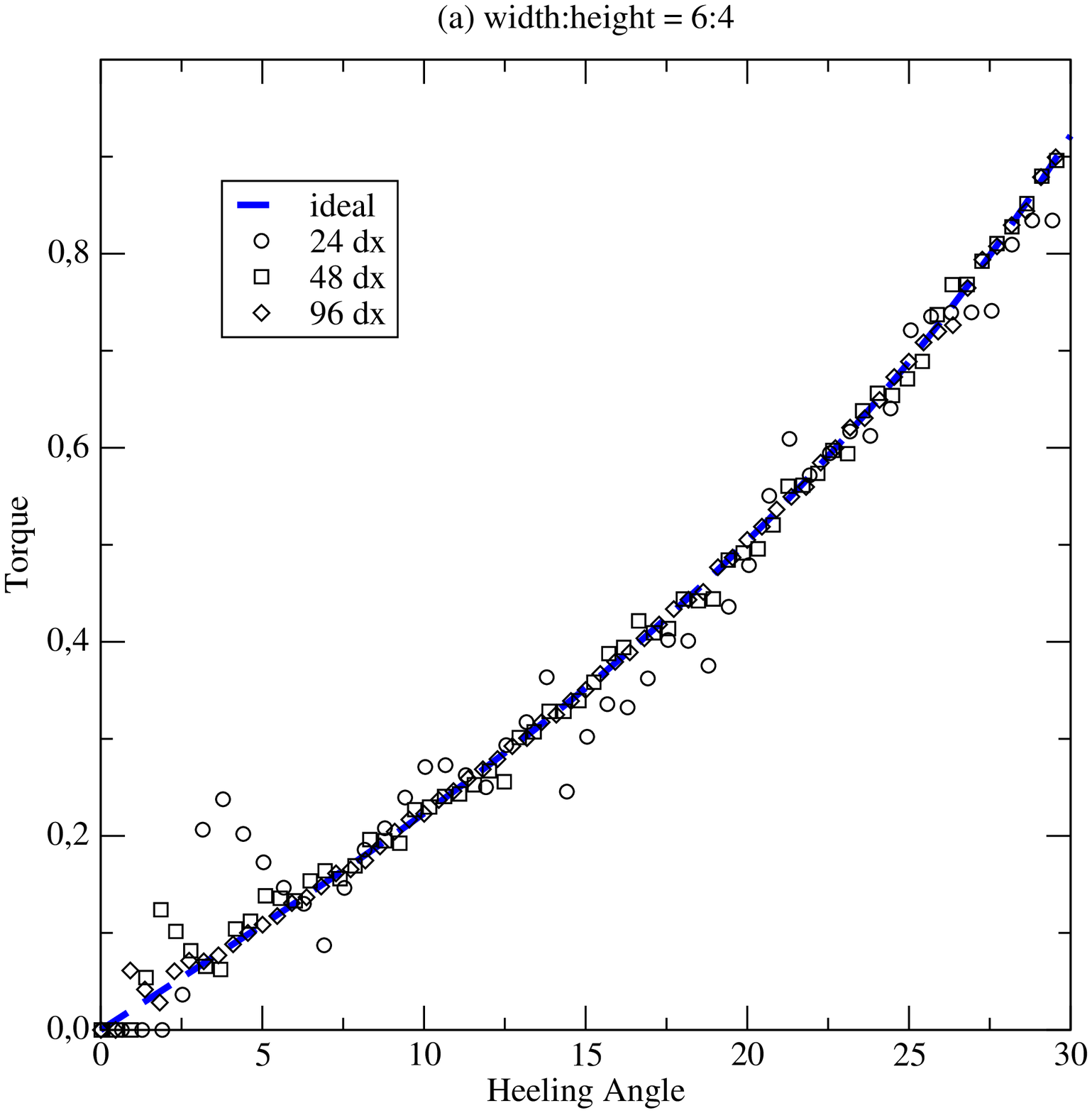} \includegraphics[scale=0.35]{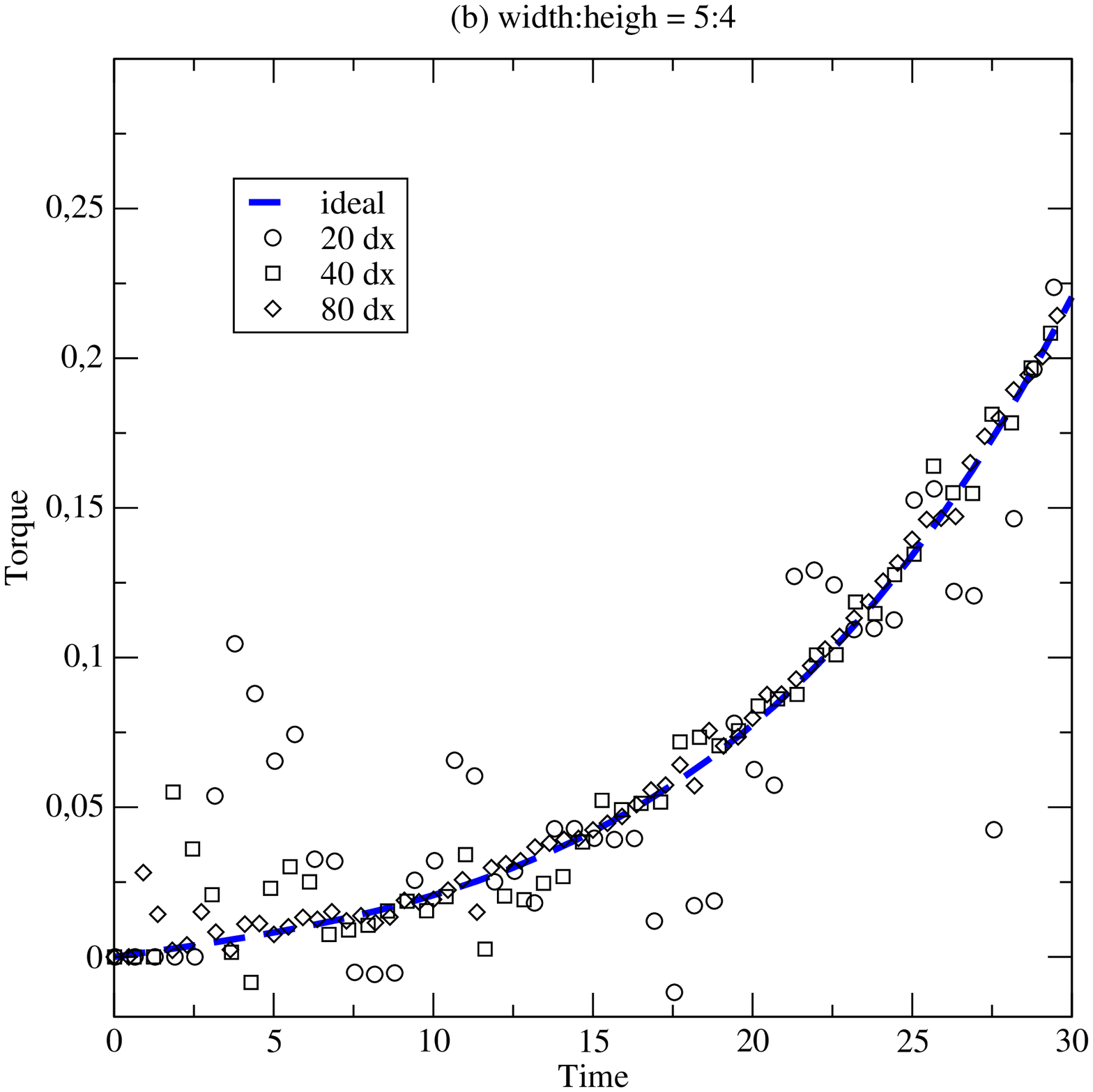}
\end{figure}

To further test the force calculation on floating bodies, the righting stability moment of
a floating cube at different angles of heel was measured in lattice Boltzmann simulations
and compared to the ideal stability curve of the structure. In all of the following cases,
the gravitational constant was chosen to be $g=10^{-4}$ and a partially filled basin was
initialized with a pressure gradient according to hydrostatic equilibrium.  The relaxation
time was $\tau=1.0$. The initial width to height ratio of the box was $6:4$ (a), and then
in a second pass a lower ratio of $5:4$ (b) was chosen. Each time, the error was examined
at three different resolutions, i.e., the box size ($b \times h$) in lattice units was $24
\times 16$, $48 \times 32$, $96 \times 64$ and $20 \times 16$, $40 \times 32$, $80 \times
64$, respectively. The length side along the axis of rotation was always chosen as $l=2
h$.  The box was positioned axis-aligned with the lattice, exactly half-immersed relative
to the free surface plane. During the simulation, the box was fixed to a constant position
and the angle of heel of the half-immersed box was varied with $\alpha$ in
$0^{\circ}..30^{\circ}$ around the longitudinal axis of the box through its center of
gravity.

Fig. \ref{fig:measuredMoment} shows the simulation results compared to the ideal stability
curve. The error was much larger in the second case (b), since the stability lever arm of
the box is shorter at the lower width per height ratio. Thus even small errors in the box
discretization have a visible influence on the behavior of the box. However, in all cases
it can be seen that the errors are decreasing at higher resolutions. Convergence towards
the ideal values was clearly given.


\section{Conclusion}
In this paper we described a method for the simulation of liquid-gas-solid flows by means
of the lattice Boltzmann method. Based on previous works, the momentum exchange method is
used to obtain fluid stresses at the surface of rigid bodies, which allows to calculate
the resulting net force and torque. A novel set of cell conversion rules
(Sec. \ref{sec:LGS}) allows unrestricted movement of the solid bodies within the domain,
including penetrations of the free surface. We demonstrated the consistency of the method
in a free advection test with a particle following a homogeneous free surface channel flow
under the influence of gravity in Sec. \ref{sec:fixedbodies}. 

For further validation of force and torque calculations on non-spherical rigid bodies, we
applied the method to the classic mechanical problem of floating stability of immersed
structures. Basic convergence towards the equlibrium state was successfully checked for
the case of box objects with square sides. We took advantage of the fundamental stability
formulae for wall-sided structures and validated the righting moment of box shaped objects
under varying angles heel. The results were found to converge  towards the ideal
values with increased spatial resolution.

As a next step it would be interesting to move on from these hydrostatic stability
examples towards dynamic ones, that include coupled interaction of liquid and solid
objects. We will present further results together with method improvements in future
publications.

\section{Acknowledgements}
For the simulations presented in this paper, the \emph{waLBerla} lattice Boltzmann
framework \cite{walberla} and the \emph{pe} rigid body phisics engine \cite{pe} have been
used. These software projects are a collaborative effort of the Chair for System
Simulation at the University of Erlangen-N{\"u}rnberg.

The author would like to thank Jan G{\"o}tz, Dr. Klaus Iglberger, Daniela Anderl,
Dr. Stefan Donath, Matthias Markl, Tobias Preclik and Florian Schornbaum for various
discussions and corrections. Finally, many thanks go to Dr. Christian Jan{\ss}en for
his kind correspondence via electronic mail.

\bibliographystyle{plainnat} 
\bibliography{lit}

\end{document}